\RequirePackage{fix-cm}

\documentclass[twocolumn]{svjour3}          
\smartqed  
\usepackage{xcolor}
\usepackage{graphicx}

\usepackage{float}

\usepackage[autostyle=false, style=english]{csquotes}
\MakeOuterQuote{"}
\usepackage{amsmath}
\usepackage{amssymb}
\usepackage{nccmath, mathtools}

\usepackage{verbatim}
\usepackage[hyphens]{url}
\usepackage[hidelinks]{hyperref}
\hypersetup{breaklinks=true}
\usepackage[authoryear,round]{natbib}

\DeclareUnicodeCharacter{00A0}{ }
\usepackage{textcomp}
\usepackage{array, makecell}
\usepackage{booktabs}
\usepackage{hhline,colortbl}

\usepackage{url}

\smartqed  
\usepackage{graphicx}
\begin{document}

\title{Effect in the spectra of eigenvalues and dynamics of RNNs trained with Excitatory-Inhibitory constraint
}


\author{Cecilia Jarne         \and Mariano Caruso 
}


\institute{Cecilia Jarne \at
              Departmento de Ciencia y Tecnolog\'ia de la Universidad Nacional de Quilmes - CONICET \\
              \email{cecilia.jarne@unq.edu.ar}           
           \and
           Mariano Caruso \at
             Fundación I+D del Software Libre, FIDESOL and Facultad de Ciencias, Universidad de Granada, España
}

\date{Received: date / Accepted: date}

\maketitle

\begin{abstract}

In order to comprehend and enhance models that describes various brain regions {it} is important to study the dynamics of trained recurrent neural networks. Including Dale\textquotesingle s law in such models usually presents several challenges. However, this is an important aspect that allows computational models to better capture the characteristics of the brain. Here we present a framework {to train networks using such constraint}. Then we have used it to train them in simple decision making tasks. We characterized the eigenvalue distributions of the recurrent weight matrices of such networks. Interestingly, we discovered that the non-dominant eigenvalues of the recurrent weight matrix are distributed in a circle with a radius less than 1 for those whose initial condition before training was random normal and in a ring for those whose initial condition was random orthogonal. In both cases, the radius does not depend on the fraction of excitatory and inhibitory units nor the size of the network. Diminution of the radius, compared to networks trained without the constraint, has implications on the activity and dynamics that we discussed here.

\end{abstract}

\section{Introduction}\label{INTRODUCCION}

One popular approach to understanding the dynamical and computational principles of population responses of neurons in the brain is based on the analysis of artificial recurrent neural networks (RNNs) trained to perform tasks inspired by behaving animals (\cite{BARAK20171, Mante2013, RUSSO2020745}). In general, the optimization of network parameters specifies the desired output but not how the output is achieved. In such a way, trained networks also serve as a source of mechanistic hypotheses of different kinds for computational neuroscience and data analyses that link neural computation to behaviour (\cite{10.1371/journal.pcbi.1004792}).

Simple models of recurrent neural networks have been successfully used to explain the different mechanisms in {the cortex, such as} decision-making, motor control, or working memory (\cite{Murphy2009, Mante2013, Zhang2021, Freedman2006, Roitman9475}). Models can be dissected to better understand how certain biological computations are mechanically implemented (\cite{DBLP:journals/neco/SussilloB13,doi:10.1146/annurev-neuro-092619-094115}).

RNNs of rate units describe biological circuits as a set of firing rates interacting through connection weights. These systems interpolate between biophysically detailed spiking-neuron models and a wider class of continuous-time dynamical systems where the units of an RNN can be interpreted as the temporal average of one or more spiking neurons (\cite{10.1371/journal.pcbi.1004792}). 

These models are used because it is well known that a nonlinear dynamical system, such as the brain, can be approximated by an RNN with a sufficient number of units (\cite{DBLP:journals/nn/Funahashi89,DBLP:journals/nn/FunahashiN93}).

One of the aspects that sometimes are omitted when {considering models of trained networks, in Computational Neuroscience, it} is the fact that neurons present differences between excitatory and inhibitory units (\cite{doi:10.1177/003591573502800330}). {Some examples of models without neuron differences describing behaviour in the motor cortex can be found in \cite{Churchland2012, Mante2013, Sussillo2015}}. It is known that in this brain area, each neuron has a fixed set of neurotransmitters, which results in either excitatory or inhibitory downstream effects. In addition to the difference in the immediate impacts on post-synaptic neurons, E and I neurons differ in other important ways, such as there are several times (4-10x) more E neurons than I neurons in the cortex (\cite{Cornford2020.11.02.364968}).

{It is important to mention that some neuronal circuits, in fact, do not obey Dale's law as it is well described in \cite{Ludwig2006, 10.3389/fncir.2018.00117}.}

{Dale's law has functional implications in sensory processing and decision-making tasks, and it plays a key role in the understanding of the structure-function relationship in the brain. In a recent work, \cite{10.3389/fnins.2022.801847} discussed and studied the functional implications when considering neural circuits that obey Dale's principle and those that violate it. The authors hypothesize that networks violating Dale's law are capable of more flexible computations, particularly for relatively small systems, and that functional or energetic constraints in large-scale systems may have promoted Dale's principle among neurons.}

Training networks with {sign-constrained} weights presents some technical challenges (\cite{10.1371/journal.pone.0220547, 10.1162/neco.2008.07-06-295}). Batch approaches for learning can handle {sign} constraints quite efficiently, but batch training of recurrent networks often leads to instabilities during the process. Some recent works have managed to incorporate this feature (\cite{ 10.1371/journal.pcbi.1004792,  10.1371/journal.pone.0220547, 10.1371/journal.pcbi.1009271, 10.1162/neco_a_01418}). 

Here a framework for gradient descent-based training of excitatory-inhibitory RNNs is presented that incorporates the biological constraints described above. An implementation is provided based on the machine learning library Tensorflow and Keras (\cite{chollet2015keras, tensorflow2015-whitepaper}). This framework is tested by applying it to different decision making tasks. The main key in present work is that we now {ask} what happens with the network\textquotesingle s activity when {sign-constraints} are successfully applied during training comparing with the same tasks {considered} in networks unconstrained.

The kind of tasks selected in this work correspond to simple decision-making tasks with temporal stimuli. The main objective was to characterize the dynamics of the recurrent networks trained in these tasks when Dale\textquotesingle s law is considered. Previously, the eigenvalues of the recurrent weight matrix of trained networks in temporal tasks have been studied without considering such constraint, and the dynamics have been characterized successfully in \cite{Jarne2022}. Now by applying {sign-constraints} during training, we obtained networks obeying Dale\textquotesingle principle. We studied the connectivity matrices and the eigenvalue\textquotesingle s  distributions of such networks.

The dynamics of trained RNNs including {Dale\textquotesingle s} law is ruled by a few dominant eigenvalues related to the task as it was observed in networks with non-differentiated units (\cite{jarne2019detailed, Jarne_2021, Jarne2022}).
We found a new effect which is the reduction of the radius on the distribution of the non-dominant eigenvalues of the recurrent weight matrices. This effect is the first time it has been reported in trained networks and does not depend on the relationship between the number of excitatory and inhibitory neurons rather the reduction depends on the initial condition.

We characterized such effect and discussed in-depth the dynamics obtained in these networks. 

The rest of the work is organized as follows. In Section \ref{MODELO}, the model and implementation details are displayed. Section \ref{resuls} shows the results and analysis carried out on the simulations. In Section \ref{disusion}, the results obtained are discussed, and finally, in Section \ref{conclusions}, the conclusions and further work are presented.

\section{Methods} \label{MODELO}

\subsection{Network Model} \label{net_2}

The dynamics of the RNN model of $N$ units is described in terms of the activity column vector function $\pmb{h}{=}(h_1,\cdots,h_N)^{\mathfrak{t}}$, where $\mathfrak{t}$ {represents} the matrix transposition. The $i-$activity component $h_i$, where $i=1,\;\cdots,N$ satisfies the following differential equation as a function of time $t$  (\cite{Hopfield3088}):

\begin{equation}
\frac{dh_i(t)}{dt}=-\frac{h_i(t)}{\tau}+\sigma \left( \sum_{j=1}^{N}W^{\mathtt{rec}}_{ij}h_j(t) +\sum_{k=1}^M W^\mathtt{in}_{ik} x_k(t) +b_{i}\right)
\label{eq-01}
\end{equation} 

where $\tau$ represent a characteristic time of the system and $\sigma$ is a non-linear activation function. The elements $W^{\mathtt{rec}}_{ij}$ are the synaptic connection strengths of to the \texttt{recurrent} weight matrix $\pmb{W}^{\mathtt{rec}}{\in}\mathbb{R}^{N\times N}$ and $x_k$ are the {components} of the column vector function of input signal $\pmb{x}{=}(x_1,\;\cdots,x_M)^{\mathfrak{t}}$. The elements $W^\mathtt{in}_{ik}$ conform the  \texttt{input} weight matrix $\pmb{W}^\mathtt{in}{\in}\mathbb{R}^{N\times M}$ which connects the input signal $\pmb{x}$ to 
each of $N$ units with activity vector $\pmb{h}$. Finally,  $\pmb{b}{=}(b_1,\;\cdots,b_N)^{\mathfrak{t}}$ is a column vector which represents the bias term. 

The network is fully connected, and matrices have recurrent weights given from a normal distribution with zero mean and variance $\frac{1}{N}$. Additional details are explained in Section \ref{proto}.

The readout of the network is given by 
\begin{equation}
z(t)= \sum_{i=1}^N W^\mathtt{out}_{i}h_i(t),
\label{eq-02}
\end{equation}
in terms of the \texttt{output} weight matrix, which in this work is a row vector
\begin{equation}
\pmb{W}^{\mathtt{out}}{=}(W^\mathtt{out}_{1},\,\cdots ,W^\mathtt{out}_{N}).
\end{equation}
Note that for binary decision-making tasks there is a unique output signal. In other words, if it is intended to distinguish between two states, this can be achieved by observing the value of a scalar quantity namely $z(t)$.

\begin{figure*}[htb!]
\begin{center}
\hspace*{-1cm}\includegraphics[width=15.5cm]{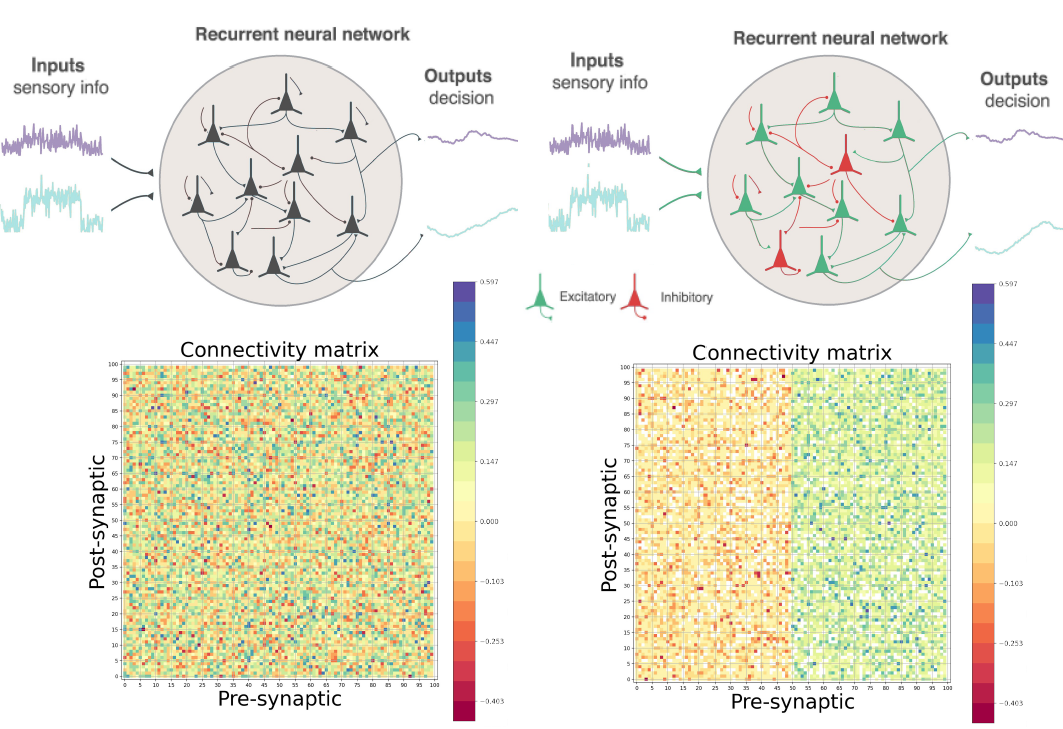}
\caption{Upper Panels shows a representation of a trained RNN without Dale\textquotesingle s constraint {(left side)} and the other of excitatory and inhibitory rate units {(right side)}. The networks receives time-varying inputs $\pmb{x}(t)$ and produces the desired time-varying outputs $z(t)$. Inputs encode task-relevant sensory information (or internal rules), while outputs indicate a decision in the form of an abstract decision variable.  $ \pmb{W}^{\mathtt{rec}}$ matrices of each network are shown in the bottom panel. Output connections are columns \textit{pre-synaptic} and input connections are rows \textit{post-synaptic}. Their values are represented by the different colors in the scale.
 \label{fig_01}}
\end{center}
\end{figure*}

For this study we considered the hyperbolic tangent, $\mathtt{tanh}$, as an activation function $\sigma$ and the characteristic time $\tau$ is set to  $1\, ms$. This nonlinear function was chosen following models used in experimental works such as \cite{RUSSO2020745, WILLIAMS20181099, Remington2018}, but it could be changed without loss of generality, as well as the time scale. A {discrete} version of the presented model through the Euler method is obtained following \cite{10.1371/journal.pcbi.1007655, 10.1371/journal.pone.0220547,Bi10530}. This is shown by the recurrence Equation \ref{eq-03}, expressed in a compact matrix form.
\begin{equation}\label{eq-03}
\pmb{h}(t_{s{+}1})=\pmb{\sigma}\pmb{\bigl(}\pmb{W}^{\mathtt{rec}}\pmb{h}(t_s)+\pmb{b} +\pmb{W}^\mathtt{in}\pmb{x}(t_s)\pmb{\bigl)},
\end{equation}
where time step is $\Delta{=}1\,ms$. The output is expressed also in a matrix form 
\begin{equation}
z(t_s)=\pmb{W}^{\mathtt{out}}\pmb{h}(t_s).
\end{equation}
A representation of the RNN model with and without Dale\textquotesingle s  principle implemented is shown in Figure \ref{fig_01}. As previously stated, there are currently two types of network work units: inhibitory (all output connections are negative) and excitatory (all output connections are positive). Respectively they are shown in the connectivity matrix on the right side of Figure \ref{fig_01}, where the output connections of $ \pmb{W}^{\mathtt{rec}}$ matrix are columns, \textit{pre-synaptic connections} and input connections are rows, \textit{post-synaptic connections}. This is shown by the different dominant colours in the scale. Cortical neurons have either purely excitatory or inhibitory effects on postsynaptic neurons.

Networks with two different excitatory-inhibitory ratios were considered: the balanced case (half of each type, as in the {studies} from \cite{10.1371/journal.pone.0220547, PhysRevLett.97.188104,10.1371/journal.pcbi.1008536}), and the case where excitatory neurons outnumber inhibitory neurons (2/3 excitatory vs. 1/3 of the inhibitory, inspired by  \cite{10.1371/journal.pcbi.1004792}).

Given that long-range projections in the mammalian cortex are excitatory exclusively (\cite{freeman2000brains}), it was assume that inputs to the network are long-range inputs from an upstream circuit, and then all elements of the input weight matrix $\pmb{W}^\mathtt{in}$ are non-negative as considered also in \cite{10.1371/journal.pcbi.1004792}.

\subsection{Training protocol}\label{proto}

Networks of 100 units were trained using backpropagation through time with the adaptive minimization method called Adam (\cite{DBLP:journals/corr/KingmaB14}). This method has been successfully used in several works for example in \cite{russo-2018} and also in our previous works (\cite{Jarne2022, Jarne_2021}).

The stimuli presented at the inputs of the networks forms the training sets. They are time series containing rectangular pulses with random noise corresponding to 10 \% of the pulse amplitude. The target output completes the set, and it will depend on which is the task parametrization, as previously reported in \cite{ Jarne2022, jarne2019detailed, Jarne_2021}. The tasks can be learned in reasonable computational time, and with good accuracy, with 100 units and 20 epochs.

Two types of initial conditions were considered for the recurrent matrices: random normal distribution and random orthogonal (\cite{chollet2015keras, jarne2019detailed}). The second case is an additional constraint, where the matrix is initialized with an orthogonal matrix obtained from the decomposition of a matrix of random numbers drawn from a normal distribution, as indicated in Section \ref{net_2}. 

Other {types} of initialization in recurrent weights has been used in different works. One example is to start with the autoconnection terms equal to 1 and the rest equal to zero (\cite{chollet2015keras, pmlr-v48-henaff16, https://doi.org/10.48550/arxiv.1504.00941}). From this approach, excellent results can be obtained for training in terms of accuracy, and this is used as a regularization technique in Machine learning. However, in this case, the weight configurations obtained are far from the biological models.

We trained 10 different networks for each initial condition, task and size. We considered three simple decision making {tasks} with temporal stimuli: \texttt{AND}, \texttt{OR} and \texttt{XOR}. As mentioned in Section \ref{INTRODUCCION}, the model was implemented in python using Keras and Tensorflow (\cite{chollet2015keras, tensorflow2015-whitepaper}). Simulations and code to perform the analysis of present work are provided in \textbf{Supplementary Information}. Table \ref{tabla1} summarizes the main parameters considered for network training. It is possible to train recurrent networks by applying Dale\textquotesingle s constraint without considering the bias term, but convergence is much more difficult for the algorithm, and we need many more training instances. The bias term also represents a neuron\textquotesingle s intrinsic minimum current value.

\begin{table}[htb!]
\centering
\begin{tabular}{!{\vrule width 1.5pt}l!{\vrule width 1.5pt}l!{\vrule width 1.5pt}}
\bottomrule[1.5pt]
\rowcolor[RGB]{175,148,145}\multicolumn{1}{!{\vrule width 1.5pt} c!{\vrule width 1.5pt}}{\textbf{parameter $\pmb{\parallel}$ condition} } & \multicolumn{1}{c!{\vrule width 1.5pt}}{\textbf{value} $\pmb{\parallel}$ \textbf{settings}}\\ 
units ($N$)                 & 100          \\
\rowcolor{gray!12}bias ($\pmb{b})$            & yes   $(\pmb{b}\not\equiv \pmb{0})$           \\
\texttt{recurrent} weight matrix      & $\pmb{W}^{\mathtt{rec}}{\in}\, \mathbb{R}^{100\times 100}$ \\
\rowcolor{gray!12}\texttt{input} weight matrix & $\pmb{W}^{\mathtt{in}}{\in}\, \mathbb{R}^{100\times 2}$ \\
\texttt{output} weight matrix               & $\pmb{W}^{\mathtt{out}}{\in}\,\mathbb{R}^{1\times 100}$    \\
\rowcolor{gray!12}Dale\textquotesingle s constraint (exc,inh)\% 
& none$-$(50,50)\%$-$(70,30)\% \\
time step ($\Delta $)                       & 1 $ms$\\
\rowcolor{gray!12} training algorithm             & \texttt{BPTT Adam} \\
initial $\pmb{W}^{\mathtt{rec}}$                         & \texttt{rand(normal$-$orthogonal)} \\
\rowcolor{gray!12} input noise        & 10\% \\
regularization                         & none  \\
\toprule[1.5pt]
\end{tabular}
\caption{Model\textquotesingle s  parameters and conditions for the network implementation and training. Code is available for implementation at \textbf{Supplementary Information}.}
\label{tabla1}
\end{table}

We implemented Dale\textquotesingle s constraint by using Keras features, which {allows} us to use any custom constraint via the function definition. This was implemented using the constraint as a subclasses of 
\href{https://www.tensorflow.org/api_docs/python/tf/keras/constraints/Constraint}{\texttt{tf.keras.constraints.Constraint()}} method. In our framework, we initially start with recurrent weight arrays initialized as random normal or random orthogonal. Then in each instance of training, it is required that the columns of the matrix, in the indicated proportion, maintain a certain sign. For specific details of how it was implemented in python see the software repository in the \textbf{Supplementary Information}.

{In general terms, the software implementation, consists of a main sofware that creates and trains the neural network. Then, a set of functions written in a library that can be called and used to control the desired contraints to implement during training: initialization, percentage of inhibitory, excitatory units, and hyperparameters in general. Some examples of such hyperparameters are the size of the network and training instances.}

{On the other hand, there is a set of independent functions that can be called to select the type of task  to train the network and its type of parameterization (task parametrization, i.e. signal noise, stimulus amplitude, among others). These libraries create the data set for training the network.}

{Our framework, which was previously used in \cite{Jarne2022, Jarne_2021}, allows us to generate the data set for training according to the task that we want to parameterize. It also allows us to perform the analysis on the connectivity of the matrix and the activity of the units.}

For this work, we have considered binary decision-making tasks, but other task such as working memory or context decision making can be used for training with the same framework.

For each network realization, the distributions of the recurrent weights were plotted and studied. The distribution moments were estimated in each case. Then, the decomposition of $ \pmb{W}^{\mathtt{rec}}$ in their eigenvectors and eigenvalues was obtained.

The realizations were studied and classified one by one. To do this, a noise-free testing set, corresponding to the input options, was used to study the response of the networks for each task. The behaviour of the units was plotted as a function of time for each of the four possible stimuli combinations.

\section{Results} \label{resuls}

The results of the numerical studies are shown in this section. The first study is on the activity of trained networks, which is shown in Section \ref{activity}. The second is on the statistics on the recurrent weight matrix and is presented in Section \ref{statistics}. The third is on the distribution of eigenvalues (Section \ref{dist_autovalores}).

\subsection{Characterizing {the activity of the units} for trained networks} \label{activity}

We {compared} the activity in trained networks with and without differentiated excitatory and inhibitory units. The case of undifferentiated units has been extensively studied in the literature. It is interesting to {note} that in these cases it is well known that different dynamic behaviours appear for the same task, such as fixed points and oscillatory states (\cite{DBLP:journals/neco/SussilloB13, BARAK20171, PhysRevResearch.3.013176, Jarne2022}).

When we trained neural networks to perform binary decision-making tasks without applying Dale\textquotesingle s law, we observed three possible states for the activity: oscillatory, fixed-point behaviour with a final constant level for each unit other than zero, and activity that decays to zero. The behaviour of such networks has been carefully described in \cite{Jarne2022}.

When we train networks applying Dale\textquotesingle s constraint, the main change in the activity of the units is that we no longer have networks where the units exhibit oscillatory behaviour. We'll go into more detail and explain why {we observe} this difference in Section \ref{disusion}. In addition, the response of the activity $\pmb{h}(t)$ of the excitatory and inhibitory units against the stimulation presents differences.

We observed that activity of the excitatory or inhibitory units is not always configured in the same way for all trained networks as it was also the case with networks with undifferentiated units. Even after applying Dale\textquotesingle s  law, we still obtained multiple configurations for the same task, as in  \cite{Jarne2022}.

\begin{figure*}[h!tb]
\begin{center}
\hspace{0cm}\includegraphics[width=14cm]{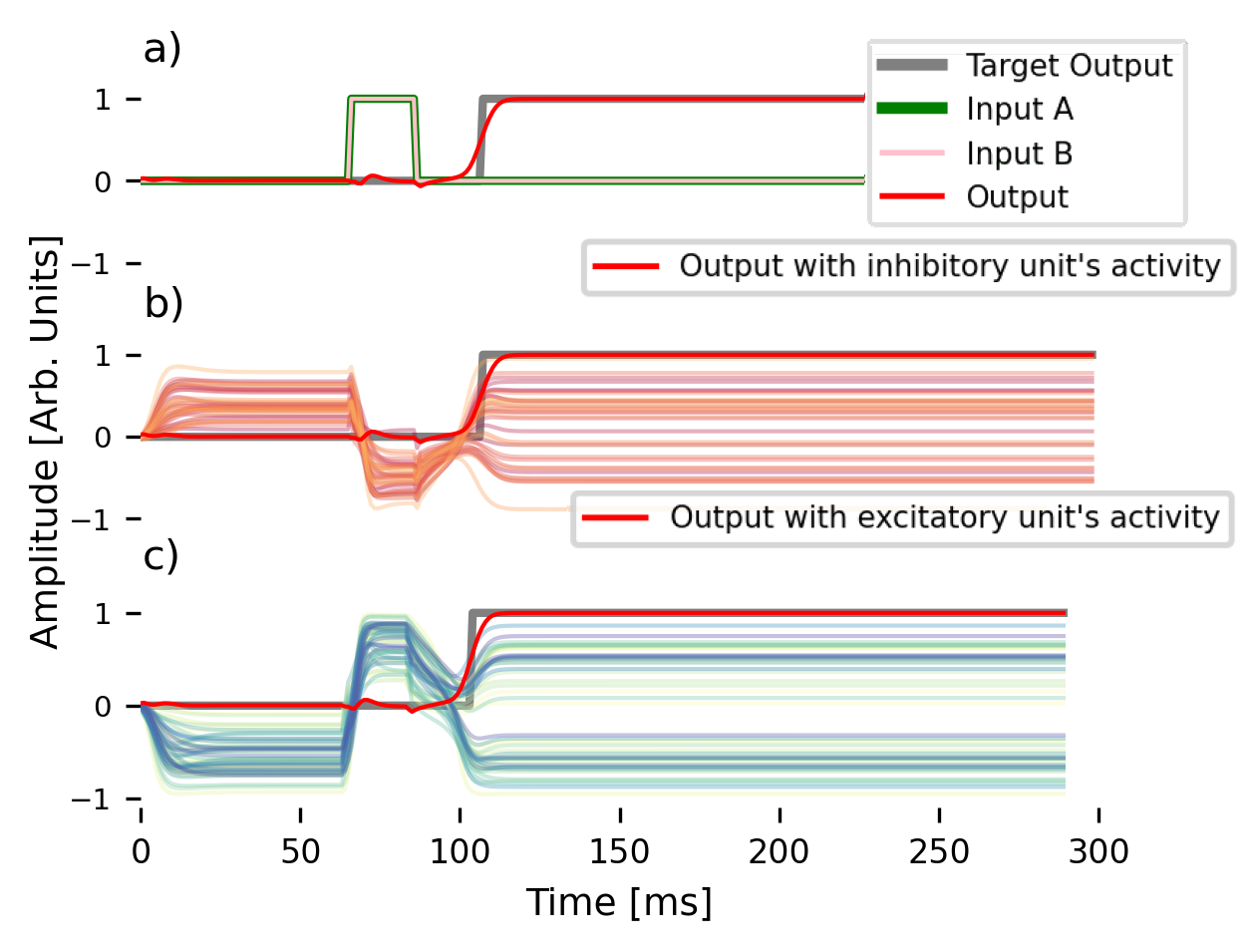}
\caption{Output signal of a RNN trained for the  \texttt{AND} task. Panel a) shows the stimuli presented simultaneously in inputs A and B (lines green and pink overlaping). The stimuli produce the trained output (red line) in agreement with the target output (gray line). Panel b) shows the activity of all the inhibitory units, and Panel c) shows all the activity of the excitatory units. Different behavior can be observed between both types of units in response to the arrival of the stimuli. The trained output is shown in all figures and is the result of the linear sum of the activity of all the units.}
\label{fig_03}
\end{center}
\end{figure*}

\begin{figure*}[htb!]
\begin{center}
\hspace{0cm}\includegraphics[width=14cm]{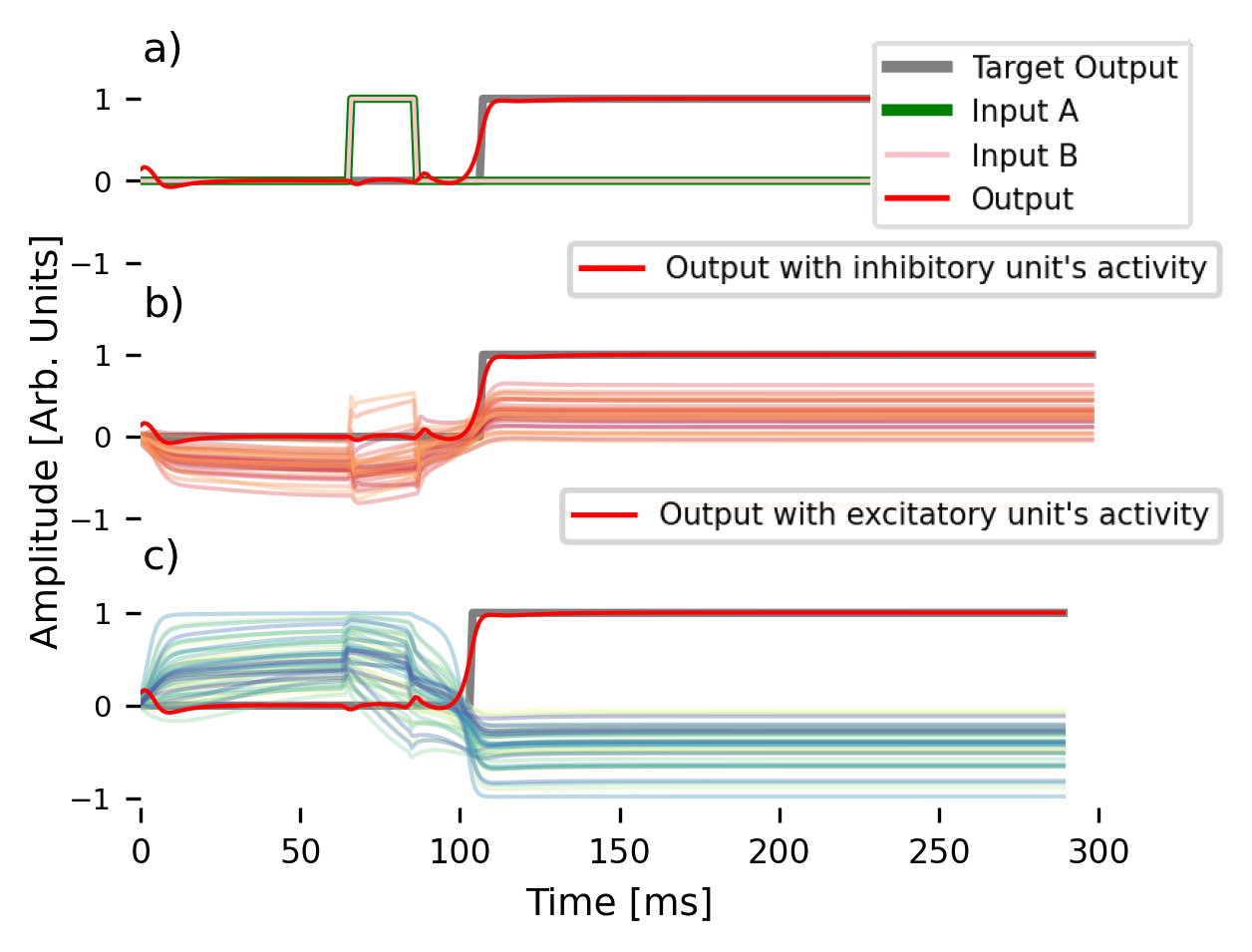}
\caption{A second example of the output signal of an RNN trained for the \texttt{AND} task. Panel a) shows the stimuli presented simultaneously in inputs A and B that produces the trained output (red line) in agreement with the target output (grey line) as expected. Panel b) shows the activity of all the inhibitory units, and Panel c) shows all the activities of the excitatory units. Again, different behaviour can be observed between both types of units in response to the arrival of the stimuli. The trained output is shown in all figures and is the result of the linear sum of the activity of all the units. The response of the excitatory/inhibitory units are different for the same task \texttt{AND} presented in Figure 2, showing that different realizations allows to obtain for the same task during training when considering the same initial configuration.}
 \label{fig_04}
\end{center}
\end{figure*}

We illustrate this behavior with an example for networks trained in the \texttt{AND} task.  Let\textquotesingle s see two different networks trained to perform the same task, as shown in Figures \ref{fig_03} and \ref{fig_04}. We compare here the high level states of the output for both realizations. Both Figures show three subplots with time series. Panel a) shows the stimuli in both inputs that, according to the training rule, will produce a high state in the output. The grey line represents the target output. The red line is the output of the trained network. The green and pink lines are the stimuli at each input. Panel b) shows the activity of all inhibitory units as a function of time and their response when receiving stimuli at the network input. Panel c) respectively shows the activity of all excitatory units. In both cases, the network responds to the behaviour described by Equations \eqref{eq-01} and \eqref{eq-02}, with the constraint that must be applied considering the 100 units, where 33 are inhibitory and the remaining 67 are excitatory.

If we observe panel b) of Figure \ref{fig_03} {and compare} it with the corresponding panel of Figure \ref{fig_04}, we see that the initial behaviour of the inhibitory units of the network is initially greater than 0. When the network receives the stimuli at the input, activity decreases and then the slope rises. After a transitory regime the final level stabilizes. While in the case of Figure \ref{fig_04}, when looking at panel b) we can see that the initial activity is less than 0 for the inhibitory units, it is disturbed when it receives the stimulus, and then it has a positive slope to finally stabilize at its final level.

We can carry out the same analysis for panel c) of both figures, whose activity for the excitatory units has very different behaviour. Despite this, in both cases, the states of the network are combined to give the result for which the output has been trained, meaning that for both inputs with a stimulus, the output must be in high-level state. The same analysis we did for the high level state of the output, can be done for the passivated state, where very different behaviour is also observed for both realizations (See Figures 1 and 2 from  \textbf{Supplementary Information}).

The above is an example of two different networks trained resulting in realizations that give different activity behaviours to produce the same output result. This is observed throughout the entire set of trained networks, with multiple types of activity for the units of networks trained for the same task (See also \textbf{Supplementary Information}).

To summarize we can say that first within the same network, the activity of excitatory and inhibitory units is differentiated in response to stimuli. Second, for the same type of task, there are multiple possible realizations and behavior of the activity where in all the realizations that we obtained, the response of excitatory and inhibitory units is differentiated. This is consistent across all binary decision-making tasks that we considered.

\subsection{On the distribution of elements of \texorpdfstring{$\pmb{W}^{\mathtt{rec}}$}{}}\label{statistics}

For a given task, the distribution of synaptic weights after training depends on a variety of factors including the initial value of the $\pmb{W}^{\mathtt{rec}}$ matrices, the choice of regularizer during training, and whether the network is tonically driven by constant external current or bias term.

\begin{figure*}[htb!]
\begin{center}
\hspace{0cm}\includegraphics[width=6.5cm]{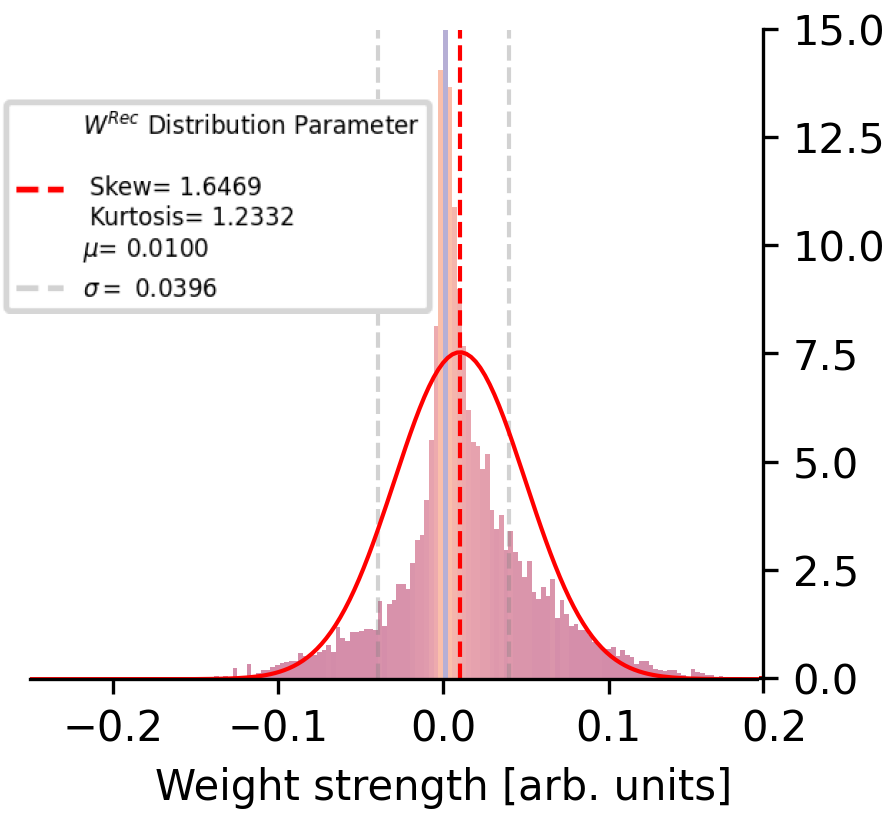}\hspace{0.5cm}
\hspace{0cm}\includegraphics[width=6.5cm]{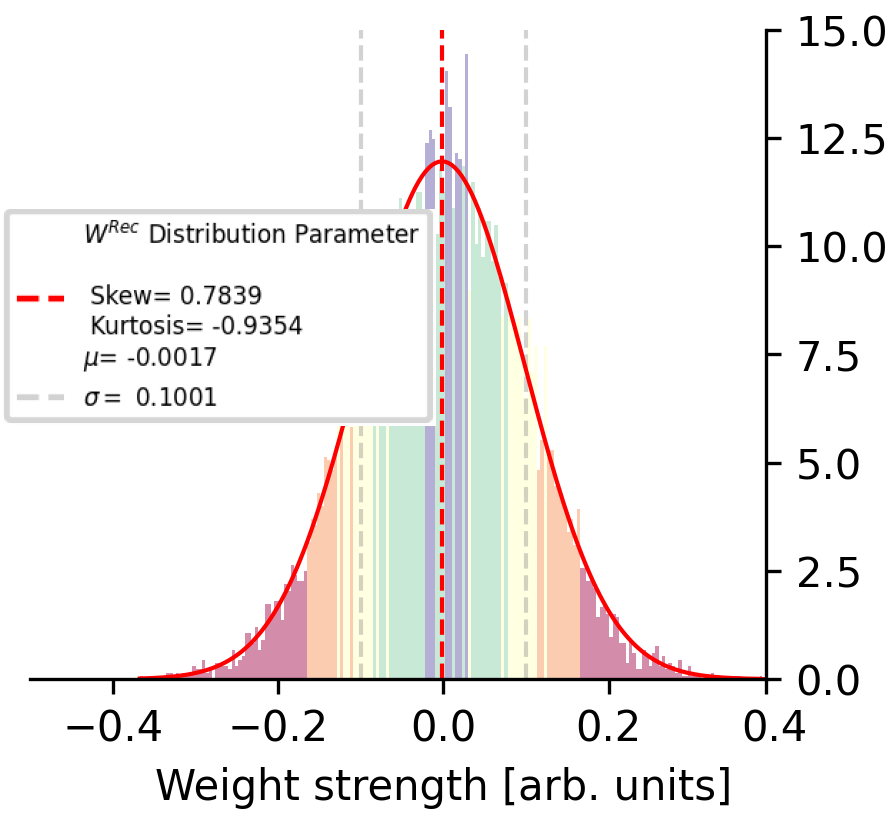}
\caption{Comparison of the weight distributions for the $\pmb{W}^{\mathtt{rec}}$ matrix of a recurrent neural network trained without the Dale\textquotesingle s law ({right} figure) versus a trained network considering such constraint ({left} figure). The second network contains $70\%$ excitatory and $30\%$ inhibitory units. It can be seen how the second distribution shrinks around zero.}
\label{fig_02}
\end{center}
\end{figure*}

The parameters of the final distribution of the recurrent weight matrices were obtained when training the networks based on the protocol described in Section \ref{proto}. The first observation is that the distributions of weights are narrower compared to those obtained without applying the Dale\textquotesingle s constraint (See Figure \ref{fig_02}). In such figure we compared two networks trained for the \texttt{AND} task, with and without the {sign-constraint}. We found that an extensive fraction of synaptic weights are set to zero by the training algorithm when the Dale\textquotesingle s constraint is applied. We also observed how the symmetry in the weight distribution is broken, as expected also for the networks with a majority of excitatory units, since it has most of its weights positive  in this case. These results, are compatible with those observed in \cite{10.1371/journal.pcbi.1004792} when Dale\textquotesingle s principle is imposed. No additional {changes} are observed in the parameters of the distribution, apart from those that account for the asymmetry.

\subsection{On the eigenvalue distribution of \texorpdfstring{$\pmb{W}^{\mathtt{rec}}$}{}} \label{dist_autovalores}

\begin{figure*}[htb!]
\begin{center}
\hspace{0cm}\includegraphics[width=6.5cm]{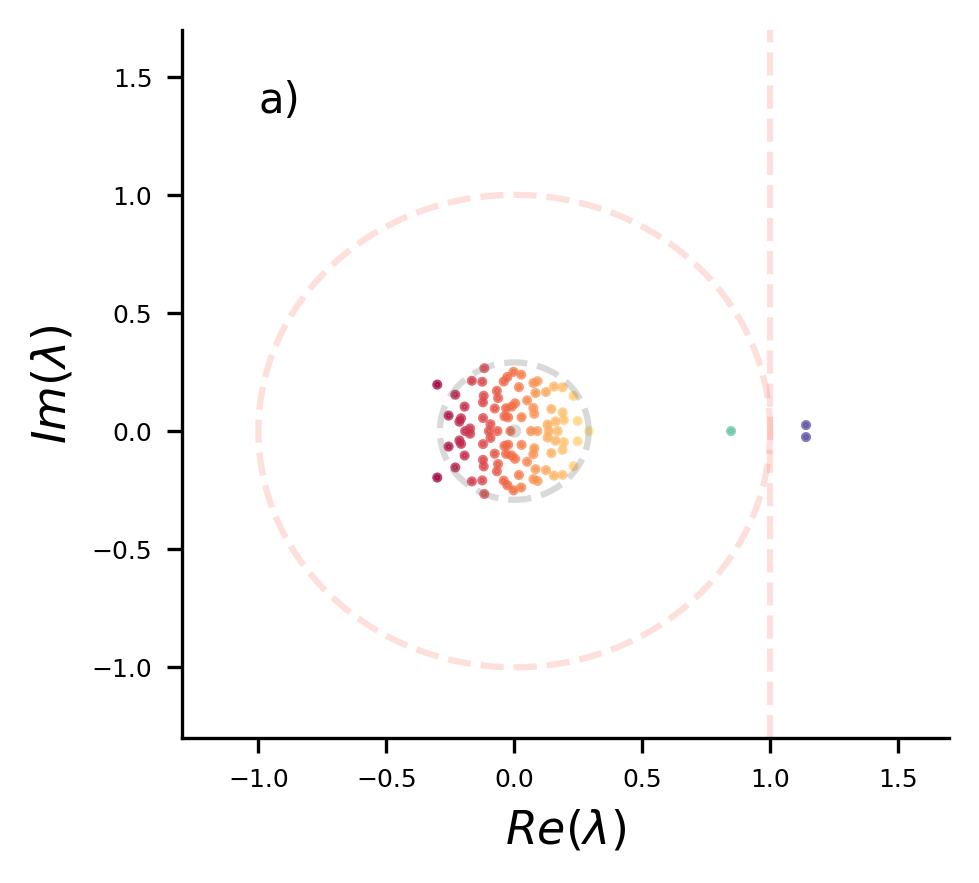}
\hspace{0cm}\includegraphics[width=6.5cm]{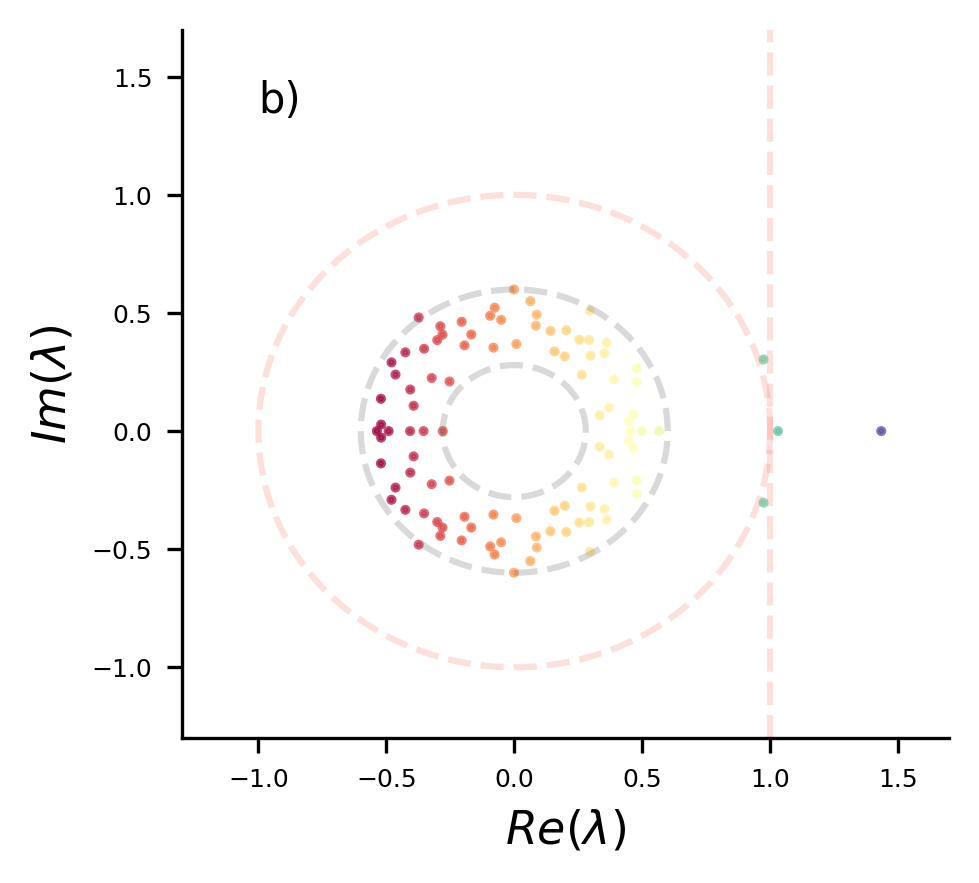}
\caption{Figure a). Distribution of eigenvalues for an RNN trained for the \texttt{AND} task with Dale\textquotesingle s constraint with $30\%$ excitatory and $70\%$ inhibitory units. The initial condition for the distribution was random normal. Again, we have a set of dominant eigenvalues associated with the modes of the trained task, whose real part is greater than unity. This is the same behavior as in networks without the Dale constraint (\cite{Jarne2022}). The novelty is that now most of the eigenvalues lie inside the circumference of radius 1 with a smaller radius. Figure b) shows the distribution of eigenvalues for a recurrent neural network with Dale\textquotesingle s constraint containing also $30\%$ excitatory and $70\%$ inhibitory units and trained for the same \texttt{AND} task from the initial condition of random orthogonal weight distribution, with the rest of the characteristics equal to the one presented in Figure a). In this case, we observe again a set of dominant eigenvalues associated with the modes of the task, but the rest lie approximately distributed in the form of a ring, again with an outer radius less than 1.}
 \label{fig_05}
\end{center}
\end{figure*}

Similar to networks without the constraint, the distributions of eigenvalues obtained when training the networks with Dale\textquotesingle s principle strongly depend on the initial condition and task. We obtained configurations where there are few dominant eigenvalues located outside the unit circle and the rest of the eigenvalues inside, but now with an inner radius less than 1. We estimated the average radius after training for the random normal initial condition and we obtained $\sim 0.29$ in average for all tasks. Figure \ref{fig_05} compares two different networks with $30\%$ of inhibitory units and $70\%$ of excitatory. The eigenvalue distribution for a trained network to perform \texttt{AND} task with Dale\textquotesingle s constraint is shown. In Figure \ref{fig_05} a) the results for the random normal condition are consistent with the observed in Rajan\textquotesingle s work for random networks (not trained), with separated inhibitory and excitatory neurons (\cite{PhysRevLett.97.188104}). The radius is reduced compared with the not differentiated units. In our case the reduction occurs also and is different. We do not have high density areas of eigenvalues. In our trained, networks the radius does not depend on the fraction of excitatory and inhibitory units.

The majority of the eigenvalues in our simulations—those that do not dominate the dynamics—are always distributed around an average radius of $\sim 0.29$, while the remaining eigenvalues—those associated with the network\textquotesingle s response to input stimuli—are farther.

\begin{figure*}[htb!]
\begin{center} 
\hspace{0cm}\includegraphics[width=10.5cm]{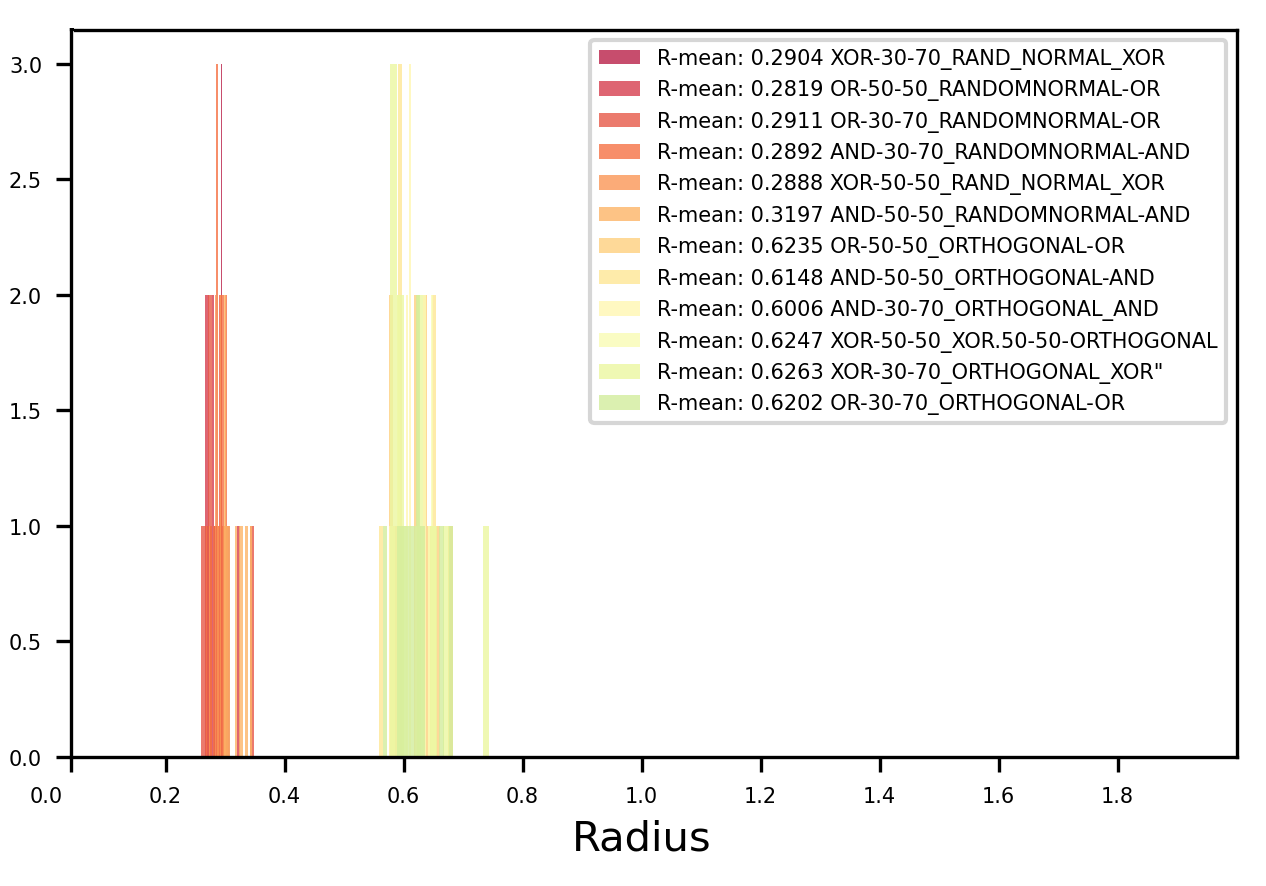}
\caption{Average of the radii for all simple decision-making tasks studied (\texttt{AND}, \texttt{OR} and \texttt{XOR}), comparing the two initial conditions considered during training random normal and random orthogonal. For each task and initial condition, 10 different networks were trained. Regardless of the percentage of excitatory or inhibitory units considered, what separates the distribution of the radii is the initial condition considered for training. The average radius after training for the random normal initial condition is $\sim 0.29$, and for the orthogonal initial condition is $\sim 0.61$, which is approximately the double.}
 \label{fig_06}
\end{center}
\end{figure*}

We found that if the initial condition for training is orthogonal in the final state, the eigenvalues are placed to form a ring with an average maximum external radius of $\sim 0.61$. An example of such trained networks is shown in \ref{fig_05} b). The minimum average radius depends on each realization. {All realizations considered in the study, for each composition and initial condition, are stored in the GitHub repository (Section \ref{suple}).} We considered all the trained networks of present work and estimated the radius in each realization for every initial condition and fraction of excitatory-inhibitory units. This study is shown in Figure \ref{fig_06}, where we can see how two groups are formed for the same type of initial condition, regardless the inhibitory-excitatory ratio. {This means that no differences between compositions are observed, only on the initial condition. The radius of non-dominant eigenvalues splits trained networks in two groups one per each initial condition.}

\begin{figure*}[htb!]
\begin{center} 
\hspace{0cm}\includegraphics[width=10.5cm]{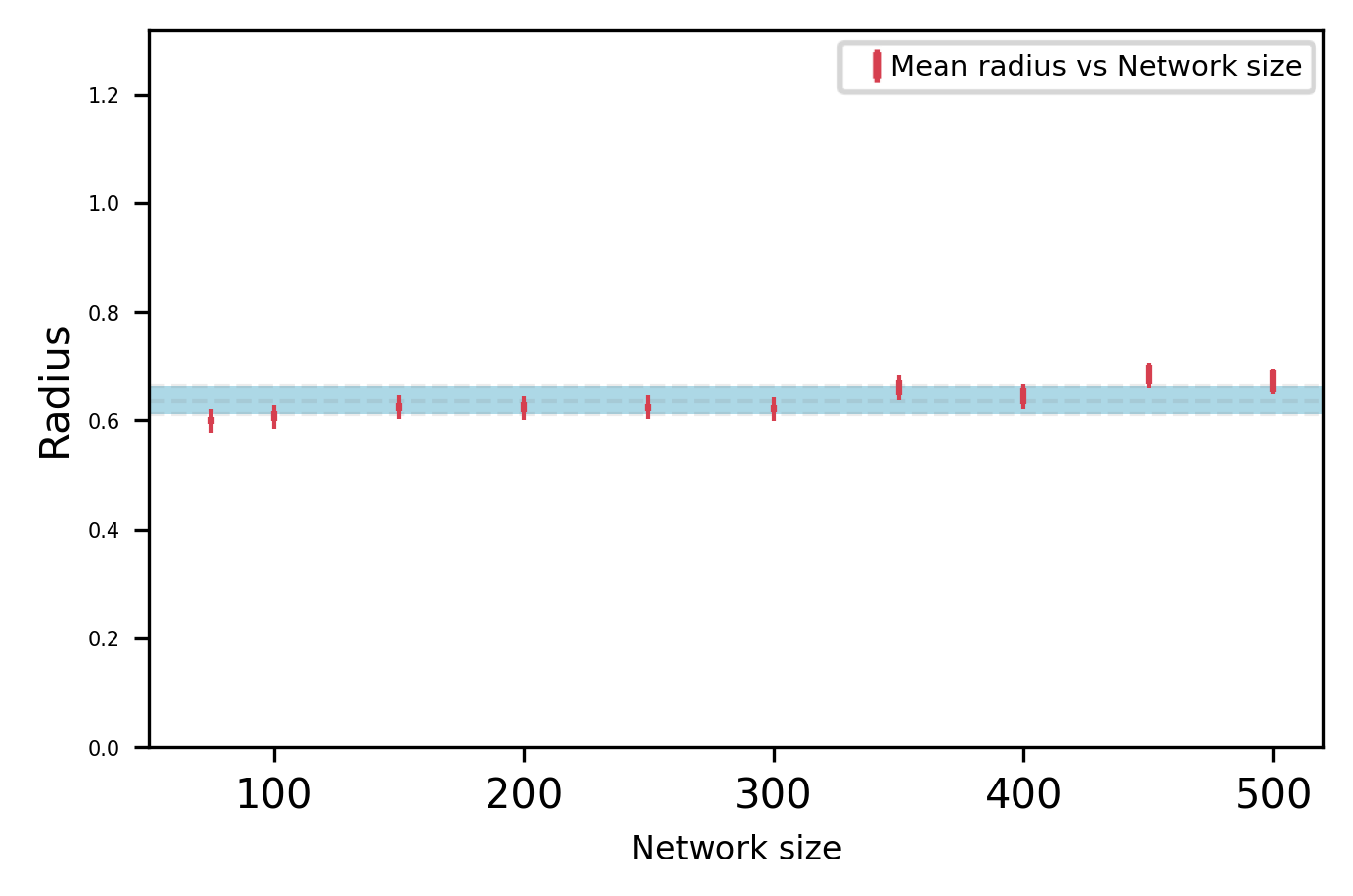}
\caption{Average radius of the distribution of internal eigenvalues for the recurrent weight matrix for the \texttt{AND} task as a function of the network size. In each network size, 10 trained  networks in said task were considered and each point is the average of them. It can be seen that the radius does not change appreciably in the range of sizes considered.}
 \label{fig_07}
\end{center}
\end{figure*}

The radius obtained from the non-dominant eigenvalues does not depend on the network\textquotesingle s size for both initial conditions. We carefully study the behaviour of the radius with the size of the network by training networks of sizes from 75 to 500 units. We trained 10 networks for the \texttt{AND} task of each network size and estimated the average radius and its uncertainties. In Figure \ref{fig_07} is shown how the radius is maintained, as the size of the network increases.

{The network size range was chosen according to the theoretical, computational and experimental studies of \cite{Churchland2012, Mante2013, Sussillo2015} and \cite{ 10.1371/journal.pcbi.1004792,  10.1371/journal.pone.0220547, 10.1371/journal.pcbi.1009271, 10.1162/neco_a_01418}. Larger neural networks, for example, of the order of 1000 units, imply a higher computational cost on training and lead to overfitting in the tasks.}

From \cite{DBLP:journals/corr/VorontsovTKP17}, it is well {known} that while orthogonal initialization of RNNs may be beneficial, maintaining hard constraints on orthogonality during training can be detrimental. The performance of models and the convergence rate of optimization can be improved by removing strict matrix orthogonality constraints. However authors observed with synthetic tasks that relaxing regularization which encourages the spectral norms of weight matrices to be close to one too much, or allowing bounds on the spectral norms of weight matrices to be too wide, can reverse these gains and may lead to unstable optimization. We followed such approach.

Requiring that the matrix maintain the signed columns (Dale's principle) with the initial conditions considered during the minimization of the gradient pushes the non-dominant eigenvalues inside the unit circle, thrusting much more inside if the initial condition was not orthogonal. This is how we obtain the distributions in small circles that we have shown in Figure \ref{fig_05} a). If the signed columns of the matrix constituted an orthogonal matrix, throughout the training instances, the eigenvalues move away from the radius equal to 1 forming a ring distribution.

The dominant eigenvalues of $\pmb{W}^{\mathtt{rec}}$ are linked to the behaviour of the activity (\cite{DBLP:journals/neco/SussilloB13, Jarne2022, doi:10.1162/neco.2009.12-07-671, PhysRevE.88.042824}). For RNNs trained in decision making tasks with not differentiated units, we observed fixed points and limit cycles in the activity linked with the dominant eigenvalues distribution been real or complex pairs. 

In the context of excitatory-inhibitory RNNs, when we have pairs of dominant eigenvalues that are complex conjugate mutually, oscillation are not observed in the activity. An example of this behavior was shown previously in Figure \ref{fig_03} and Figure \ref{fig_04} from Section \ref{activity}, and also in Figures 1 and 2 from \textbf{Supplementary Information}.

In fact, when training networks with differentiated units {using} the same method and initial conditions that we used before, we now obtain distributions of eigenvalues very similar to the previous ones, with respect to the number of dominant eigenvalues, but whose real parts are much greater than the imaginary part, producing the effect of the absence of limit cycles. The reason for the unobservability of such an oscillation is because these eigenvalues are outside the unit circle in the complex plane, which implies that the linear version of the original system has an unstable behaviour, in the sense that the signal grows indefinitely in time. Moreover, since the real part of such eigenvalues is much larger than the imaginary part in absolute value, the signal tends to grow faster or more explosively than it can tend to oscillate. In other words, if the inverse of the real part of such eigenvalues is related to a characteristic growth time of the linear system and the quotient of $2\pi$ over the absolute value of the imaginary part corresponds to the period, the remarkable inequality between these parts shows that the signal is dominated by exponential growth. 

In addition to the non-oscillation of the activity, we observe that it quickly stabilises, because the explosively increasing behaviour, which we have described above, is crushed by the activation field $\pmb{\sigma}$ which tends to  flattening  the extreme values of activity signal of the units.

\section{Discussion}\label{disusion}

The arguments presented above allows us to explain {roughly} the activity and the output signals of the trained network as a function of the initial impulses. They can be formalized mathematically and discussed in more detail. To start, we will consider the linear version of Equation \eqref{eq-01}, associated with the linearization of the system and written in matrix form

\begin{equation}\label{lin-system}
\frac{d\pmb{h}(t)}{dt}=\pmb{(}{-}\pmb{I}{+} \pmb{W}^{\mathtt{rec}}\pmb{)}\,\pmb{h}(t) +\pmb{W}^\mathtt{in} \pmb{x}(t)+\pmb{b} 
\end{equation}

note that $\pmb{I}$ is the $N{\times}N$ identity matrix. The general general solution of \eqref{lin-system} according to \cite{Polyanin, Arnold} is given by:

\begin{equation}\label{sol.lin-system}
\pmb{h}(t)=\exp{[\pmb{M}t
]}\pmb{h}(0)+\int_0^t \exp{\pmb{[M}(t-u)\pmb{]}}\pmb{(}\pmb{W}^\mathtt{in} \pmb{x}(u)+\pmb{b} \pmb{)}du
\end{equation}

where $\pmb{M}{=}{-}\pmb{I}{+} \pmb{W}^{\mathtt{rec}}$. Since the identity matrix {commutes} with any other matrix, the exponential matrix can be factorized as $\exp{[t\pmb{M}
]}{=}e^{-t}\exp[t\pmb{W}^{\mathtt{rec}}]$ (\cite{Arnold}). 

The input signals $\{x_k\}_{k\in\{1,M\}}$ are assumed to be rectangular pulses or $\Pi$ functions, as explained in Section \ref{proto}. They can be defined as: $x_k(t){:=}a_k\Pi_{[0,T_k]}(t)$, where the amplitudes $a_k$ are real numbers and $\Pi_{[0,T_k]}(t)=1$, iff $t\in [0,T_k]$ and  $\Pi_{[0,T_k]}(t)=0$, iff $t \not\in [0,T_k]$. Here $k$ is the number of inputs, which for our considered tasks is 2, but since we want to write a general solution for a general case, in the following equations, we will considered as a general index. 

The solution shown in Equation \eqref{sol.lin-system} has two terms, where the second describes a \textit{transient} regime caused by the initial abrupt change at $t{=}0$ from the input signal pulses and  depends on each time $T_k$ and each amplitude $a_k$. After a time $T{=}\max \pmb{\{} T_k{:}\, k{\in}\{1,M\}  \pmb{\}} $,  that \textit{transient} regime ends to begin the \textit{persistent} regime, described by the first term in Equation \eqref{sol.lin-system}. The general solution for Equation \eqref{sol.lin-system} can be formally expressed as 
\begin{equation}\label{sol2.lin-system}
\pmb{h}(t)=e^{-t}\exp[t\pmb{W}^{\mathtt{rec}}]\pmb{\alpha}+\pmb{\beta},
\end{equation}
the essential difference between each regime is not in a substantial change in the oscillation frequency nor in the decay times, but rather that the constants are different in each regime. Specifically, the multiplicative $\pmb{\alpha}$ and additive $\pmb{\beta}$ constants which contribute in the following two situations:  during the input signals for $t\in[0,T]$ and after these signals has elapsed $t>T$. In particular, $\pmb{\alpha}$ and $\pmb{\beta}$ are vectors that change discontinuously when times $t$ goes from $[0,T]$ to $(T,\infty)$, according to the respective state of the inputs signals being on$-$off. In particular $\pmb{\alpha}$ is defined as

\begin{equation}\label{alfa}
\pmb{\alpha}{:=}\begin{cases} 
\pmb{h}(0){+}(\pmb{W}^{\mathtt{rec}}{-}\pmb{I})^{-1}(\pmb{W}^{\mathtt{in}}\pmb{a}{+}\pmb{b}): & t{\in}[0,T],\\
\pmb{h}(0){+}(\pmb{W}^{\mathtt{rec}}{-}\pmb{I})^{-1}\pmb{b}: & t{\in}(T,\infty),
\end{cases}
\end{equation}
where $\pmb{a}$ is the column vector that contains the amplitudes of the rectangular pulses of the input signals and $\pmb{\beta}$ can be defined as  $\pmb{\beta}{:=}\pmb{h}(0)-\pmb{\alpha}$.

The eigenvalue neighborhood of $\pmb{W}^{\mathtt{rec}}$, i.e. its spectrum location in complex plane, is showed in Figure \ref{fig_05}. Note that $\pmb{W}^{\mathtt{rec}}$ has $N$  eigenvalues. As previously indicated in Section \ref{MODELO}, $N$ is also the number of network units. In particular,  $\pmb{W}^{\mathtt{rec}}$ is diagonalizable because it has pairwise distinct eigenvalues.

The final form of the solution shown in Equation \eqref{sol2.lin-system} is determined by $\exp[t\pmb{W}^{\mathtt{rec}}]$, which can be calculated using the fact that the matrix $\pmb{W}^{\mathtt{rec}}$ is diagonalizable, the case where it is non-diagonalizable is described in the \textbf{Supplementary Information}. In any case, this matrix is real, because it represents the connection weights between neurons. For a given pair of {eigenvalues} and its associated eigenvector $(\lambda,\pmb{\xi})$, $\pmb{W}^{\mathtt{rec}}\pmb{\xi}{=}\lambda\pmb{\xi}$, the complex conjugate pair $(\lambda^*,\pmb{\xi}^*)$ is also an eigenvalue and its associated eigenvector of this matrix.

Let us denote the eigenvalues of $\pmb{W}^{\mathtt{rec}}$ by $\{\lambda_i\}_{i=1,{\cdots},N}$, where $\nu$ are reals and $\mu$ are pair of complex conjugates. The computation of $\exp[t\pmb{W}^{\mathtt{rec}}]$ allows us to express the  formal solution of Equation \eqref{sol2.lin-system} as a linear combination of the eigenvector basis, following \cite{Arnold} and shown in Equation \ref{sol3.lin-system}.

\begin{equation} 
\label{sol3.lin-system}
\begin{aligned}
\pmb{h}(t)&{=}e^{-t}\bigg{\{}\sum_{k=1}^\nu c_ke^{\lambda_k t}  \pmb{\xi}_k+
\sum_{l=\nu{+}1}^{\nu{+}\mu} c_le^{\lambda_l t}  \pmb{\xi}_l+c_l^{*}e^{\lambda_l^{*} t}  \pmb{\xi}^{*}_l\bigg{\}}+\pmb{\beta}
\end{aligned}
\end{equation}

where $\lambda_l{=}\alpha_l{+}\dot{\iota}\omega_l$, if $l{\in}\{\nu{+}1,{\cdots},\nu{+}\mu\}$, and $N{=}\nu{+}2\mu$. Each term of the second sum of Equation \eqref{sol3.lin-system} is given by the real part of $c_le^{\lambda_l t}  \pmb{\xi}_l$, which is explicitly given by 
$ e^{\alpha_l t}\big{[}{\operatorname{Re}}(c_l\,\pmb{\xi}_l)\,cos(\omega_l t)-{\operatorname{Im}}(c_l\,\pmb{\xi}_l)\,sin(\omega_l t)\big{]}$. 

In this way, the solution \eqref{sol2.lin-system} is a linear combination of $e^{\lambda_k t}$ for $k{=}1,{\cdots},\nu$ and $\{e^{\alpha_l t}cos(\omega_l t),e^{\alpha_l t}sin(\omega_l t)\}$ for $l{=}\nu{+}1,{\cdots},\nu{+}\mu$ given by 

\begin{equation}
\label{soluc.lin-system}
\begin{split}
\pmb{h}(t){=}e^{-t}\bigg{\{}\sum_{k=1}^\nu c_ke^{\lambda_k t}  \pmb{\xi}_k{+} \\
\sum_{l=\nu{+}1}^{\nu{+}\mu} e^{\alpha_l t}\big{[}{\operatorname{Re}}(c_l\,\pmb{\xi}_l)\,cos(\omega_l t){-}{\operatorname{Im}}(c_l\,\pmb{\xi}_l)\,sin(\omega_l t)\big{]}\bigg{\}}{+}\pmb{\beta}
\end{split}
\end{equation}

The contribution of each term of the first sum of Equation  \eqref{soluc.lin-system} to the activity $\pmb{h}(t)$ is such that the real eigenvalues $\lambda_k$ are linked to a characteristic time of exponential growth $\alpha_k{>}0$ (or decay $\alpha_k{<}0$). On the other hand, the contribution of each term of the second sum of Equation \eqref{sol2.lin-system} to the activity $\pmb{h}(t)$ is such that the real part of the complex eigenvalues $\lambda_l{=}\alpha_l{\pm}\dot{\iota}\omega_l$ are linked to a characteristic time of the exponential growth $\alpha_l{>}0$ (or decay $\alpha_l{<}0$), while its imaginary part $\omega_l$ contributes to the period of an oscillating behaviour given by sine and cosine functions. 

We can relate the constants $\{c_i\}_{i=1,{\cdots},N}$ in the linear combination shown in Equation \eqref{sol3.lin-system}, with $\pmb{\alpha}$ from Equation \eqref{sol2.lin-system}. According to expression \eqref{alfa}, it can be written also depending on the initial activity, the \texttt{recurrent} weight matrix,  the \texttt{input} weight matrix, the amplitudes of the \texttt{input} signal and the bias.

Because $\pmb{W}^{\mathtt{rec}}$ is diagonalized, we have a basis of eigenvectors that can be arranged as columns to form a non-singular matrix  $\pmb{S}{=}(\pmb{\xi}_1,{\cdots},\pmb{\xi}_N)$. Then $\pmb{W}^{\mathtt{rec}}{=}\pmb{S\Lambda S}^{-1}$ where $\pmb{\Lambda }{=}diag(\lambda_1,{\cdots},\lambda_N)$. Using this base, we can calculate $\exp[t\pmb{W}^{\mathtt{rec}}]{=}\pmb{S}\exp[t\pmb{\Lambda}]\pmb{S}^{-1}$, thus $\exp[\pmb{\Lambda}t]{=}diag ({\cdots},e^{\lambda_kt},{\cdots})$ and the product $\pmb{S}\exp[\pmb{\Lambda}t]{=}({\cdots},e^{\lambda_it}\pmb{\xi}_i,{\cdots})$. Finally the identification of a column vector $\pmb{c}$ that contains the constants $\{c_i\}_{i=1,{\cdots},N}$ is performed as:
\begin{equation}\label{c,alfa}
\pmb{c}{=}\pmb{S}^{-1}\pmb{\alpha},
\end{equation}
also using the expression \eqref{alfa}, $\pmb{c}$ can be also written in terms of $\pmb{h}(0)$, $\pmb{W}^{\mathtt{rec}}$, $\pmb{W}^{\mathtt{in}}$, $\pmb{a}$ and $\pmb{b}$.

Regarding the long-term dynamics, any eigenvalue of $\pmb{M}{=}{-}\pmb{I}{+}\pmb{W}^{\mathtt{rec}}$, whose real part is less than zero will be asymptotically stable (\cite{Arnold}). This translates to the eigenvalues of $\pmb{W}^{\mathtt{rec}}$ that lie inside the unit circle of the complex plane and are neglected. Another argument that allows us to neglect several terms in the solution shown in Equation \eqref{sol3.lin-system} is through the concept of \textit{dominants} eigenvalues of $\pmb{W}^{\mathtt{rec}}$ (\cite{GolubLoan,AntonRorres}).  Without loss of generality we can assume that its  eigenvalues are relabeled in order to distinguish the following two groups
\begin{equation}
|\lambda_1|{\geq}\cdots{\geq} |\lambda_d|{>}|\lambda_{d+1}|{\geq}\cdots{\geq} |\lambda_N|
\end{equation}
where $|z|$ is the absolute value or modulus for a complex number $z$. These dominant $d-$eigenvalues can be real or complex.

In this sense, we call the first d eigenvalues dominants. Let us write the vector $\pmb{\alpha}$ in the basis of eigenvectors $\{\pmb{\xi}_i\}_{i=1,{\cdots},N}$ as $\pmb{\alpha}{=}\sum_{i=1}^N c_i \pmb{\xi}_i$, which is exactly follow from Equation \eqref{c,alfa}, in order to compute the product 
\begin{equation}\label{iteration}
\begin{aligned}
(\pmb{W}^{\mathtt{rec}})^n\pmb{\alpha}&=\sum_{i=1}^N c_i\lambda_i^n\pmb{\xi}_i\\
&=\bigg{[}\sum_{i=1}^d c_i\left(\frac{\lambda_i}{\lambda_1}\right)^n\pmb{\xi}_i+\sum_{i=d+1}^N c_i\left(\frac{\lambda_i}{\lambda_1}\right)^n\pmb{\xi}_i\bigg{]}\lambda_1^n.
\\
\end{aligned}
\end{equation}
Increasing $n$ in the second sum of expression \eqref{iteration} which starts from 0. In this way $\exp[t\pmb{W}^{\mathtt{rec}}]\pmb{\alpha}{\simeq} \sum_{i=1}^d {c_ie^{\lambda_i t}\pmb{\xi}_i}$. If a given complex eigenvalue is dominant, then its complex conjugate will be also dominant, since both have the same modulus. Only the dominant eigenvalues (real or complex) contributes to the activity signal $\pmb{h}$ 
\begin{equation} 
\label{sol.dom.lin-system}
\pmb{h}(t){\simeq}e^{-t}\sum_{i=1}^d c_ie^{\lambda_i t}  \pmb{\xi}_i+\pmb{\beta}
\end{equation}
Any dominant eigenvalue of $\pmb{W}^{\mathtt{rec}}$ such that its real part is greater than $1$ will contribute with an exponential growth to the activity associated to the linear version of the RNN. But it should not be lost sight of exactly that: that analysis was performed on the linear version of the RNN, where the activation function $\pmb{\sigma}$ in Equation \eqref{eq-01} would saturate any unbounded growth in the activity associated with the RNN. 

Although there are oscillating functions that contribute to $\pmb{h}$ in Equation \eqref{sol3.lin-system}, no effective oscillations are observed in the activities reported in Figures \ref{fig_03} and \ref{fig_04}. This is because the complex pair $(\lambda,\lambda^*)$ of eigenvalues of \texttt{recurrent} matrix, shown in Figure \ref{fig_05}, are such that $|{\operatorname{Re}(\lambda)}|{\gg} |{\operatorname{Im}(\lambda)}|$, and the exponential, growth or decay, behavior dominates the scene versus the oscillating behavior. 


One of the aims is to train RNNs not simply to maximize the network’s performance, but to train networks so that their performance matches that of behaving animals, while both network activity and architecture are as close to biology as possible. The incorporation of separate excitatory and inhibitory populations and the ability to constrain their connectivity is an important step towars that. One of the main contributions of present work is providing a simple framework to apply such constraint and study different decision-making tasks. Using such framework, we trained networks with composition $(50,50)\%$ and $(30,70)\%$, like the ones we have described before in \cite{Jarne2022}. We observed that networks with balanced composition are more easy to train than mostly excitatory networks. In this second case, more instances of training are necessary. There is a competition between maintaining the constraint and minimizing the cost function despite this, in these cases.

Again, when appling Dale\textquotesingle s law, different realizations for the same task were obtained with different dynamical behaviours, and the trained networks are generally non-normal.

Also, it is interesting to note how, compared to networks trained unconstrained, the number of dominant eigenvalues associated with the modes of the task remains the same. However, given the contraction of the radius of the {remaining} eigenvalues, the activity after the disturbance does not remain oscillating, as explained above. One may think that the radius depended on the fraction of inhibitory and excitatory units, similarly like in random networks, but this is not the case for trained networks (\cite{PhysRevLett.97.188104}).

If this were the regimen of the brain, it would not be possible to obtain oscillations in activity as the results observed in EEG experiments (\cite{freeman2000brains}). We have to rethink a different way to configure RNNs during training, if oscillatory activity is part of the model describing a particular brain area. 

Comparing to previus works, our model, is different from \cite{10.1371/journal.pone.0220547} in several aspects. First, the balance is parametric in this model in the sense described by \cite{10.1371/journal.pone.0220547}, which is different from the dynamic balance they also described. Second, the input weights are trained, as well as all the recurrent and output weights. Our model has in common with \cite{10.1371/journal.pone.0220547} that the weight distributions present many zeros in the connectivity values after being trained, and that most of the non-dominant eigenvalues are grouped in a circle. But in our case, given the initial conditions where the radius is 1 or less than 1, the final distributions obtained have a smaller radius, producing the attenuation of the activity.

There are other more complex tasks that can be considered, such as the so-called Working Memory tasks. Some of them are, for example, Flip Flop and Parametric Working Memory (\cite{Freedman2006, Roitman9475}). There are also binary decision tasks that are more complex, for example, depending on a context, some can be context-dependent decision-making, delay match to sample or contextual multitasking (\cite{Murphy2009, Mante2013, Zhang2021}). We have been able to train excitatory and inhibitory constrained networks to perform these tasks, but characterizing the activity in these cases is more complex will be analyzed in further works.

\section{Conclusions}\label{conclusions}

In present work, a framework for gradient descent-based training of excitatory-inhibitory RNNs has been described and demonstrated the application of such framework to tasks which are inspired by experimental paradigms in systems neuroscience.  We have discovered, and described, how the non-dominant eigenvalues of the recurrent weight matrix are distributed in a circle with radius less than 1 for those whose initial condition prior to training was random normal and in a ring for those whose initial condition was random orthogonal. In both cases the radius does not depend on the fraction of excitatory and inhibitory units nor on the size of the network, \textcolor{red}{for the considered range}. It depends on the initial condition for training. Diminution in the radius of the non-dominant eigenvalues, compared to networks trained without the constraint, has implications for unit activity attenuating the expected oscillations which have been explained in mathematical terms. Despite the ease of the task, our computer simulations offer a framework for investigating the stimulus-driven dynamics of biological neural networks and further inspire theoretical research.

Future work will include the study of more complex tasks such as working memory tasks or contextual decision making.

\section{Supplementary Information}\label{suple}

The \texttt{code} of this analysis are available at the following Github repository

\begin{center}
\href{https://github.com/katejarne/Excitatory-inhibitory}{\textcolor{blue}{\texttt{https://github.com/katejarne/Excitatory-inhibitory}}}
\end{center}

Supplementary information in pdf includes additional examples for all tasks and initial conditions.

\section*{Acknowledgments}

The present work was supported by CONICET and UNQ. Authors acknowledge support from PICT 2020-01413.

\section*{Conflict of interest}

The authors declare that they have no conflict of interest.


\bibliographystyle{spbasic}      
\bibliography{mybibfile.bib}

\end{document}